\documentclass[aps,prl,notitlepage,amsmath,amssymb,amsfonts,superscriptaddress,twocolumn]{revtex4-2}

\makeatletter
\def\@hangfrom@section#1#2#3{\normalsize\@hangfrom{#1#2}#3}%\MakeTextUppercase{#3}}%
\def\@hangfroms@section#1#2{\normalsize#1#2}%\MakeTextUppercase{#2}}%
\makeatother

\usepackage{graphicx}
\usepackage{float}
\usepackage{bm}
\usepackage{bbm}
\usepackage[dvipsnames,x11names]{xcolor}
\usepackage[colorlinks=true,citecolor=blue,linkcolor=blue,urlcolor=RoyalBlue]{hyperref}
\usepackage{orcidlink}
\usepackage[percent]{overpic}
\usepackage{tabularx}
\usepackage{multirow}
\usepackage{CJK}

\newcommand{\bluefour}[1]{\textcolor{Blue4}{#1}}
\newcommand{\panelgraphics}[5][]{%
  \begin{overpic}[#1]{#2}
    \put(#4,#5){\textbf{#3}}
  \end{overpic}%
}

\newcommand{\HOne}{H_{1}^{thr}}
\newcommand{\HTwo}{H_{2}^{thr}}
\newcommand{\Hz}{H_z^{\ast}}
\newcommand{\Hx}{H_x^{\ast}}

\newcommand{\Hchi}{H_{\chi}^{\ast}}
\newcommand{\Hgap}{H_{\Delta}^{\ast}}
\newcommand{\HN}{H_{\rm N}}
\newcommand{\Heff}{\mathcal{H}_{\rm eff}}
\newcommand{\Mphi}{M_{\phi}}
\newcommand{\MphiZero}{M_{\phi}^{(0)}}
\newcommand{\Mxe}{M_x^{\rm eff}}
\newcommand{\Mye}{M_y^{\rm eff}}
\newcommand{\Mze}{M_z^{\rm eff}}
\newcommand{\chiphi}{\chi_{\phi}}
\newcommand{\chiphiZero}{\chi_{\phi}^{(0)}}
\newcommand{\msz}{m_z^{\rm stag}}

\begin{document}
\begin{CJK*}{UTF8}{gbsn}

\title{Anomalous magnetocaloric effects in the quasi-one-dimensional antiferromagnet BaCo$_2$V$_2$O$_8$}

\author{Jiahao~Yang~(杨家豪)~\orcidlink{0000-0001-7670-2218}}
\thanks{These authors contributed equally to this work.}
\affiliation{International Center for Quantum Materials, School of Physics, Peking University, Beijing 100871, China}
\affiliation{Beijing Key Laboratory of Quantum Devices, Peking University, Beijing 100871, China}

\author{Chao Dong~(董超)}   % dongchao@hust.edu.cn
\thanks{These authors contributed equally to this work.}
\affiliation{Wuhan National High Magnetic Field Center and School of Physics, Huazhong University of Science and Technology, Wuhan 430074, China}

\author{Xinlong Shi~(史昕龙)}   % d202280125@hust.edu.cn
\thanks{These authors contributed equally to this work.}
\affiliation{Wuhan National High Magnetic Field Center and School of Physics, Huazhong University of Science and Technology, Wuhan 430074, China}

\author{Zhuo Wang~(王卓)}  % D202380138@hust.edu.cn
\affiliation{Wuhan National High Magnetic Field Center and School of Physics, Huazhong University of Science and Technology, Wuhan 430074, China}

\author{Tiantian Li~(李田恬)}   % tiantiali9-c@my.cityu.edu.hk
\affiliation{State Key Laboratory of Quantum Functional Materials and Department of Physics, Southern University of Science and Technology, Shenzhen 518055, China}

\author{Liusuo Wu~(吴留锁)}     % wuls@sustech.edu.cn
\affiliation{State Key Laboratory of Quantum Functional Materials and Department of Physics, Southern University of Science and Technology, Shenzhen 518055, China}

\author{Junfeng Wang~(王俊峰)}  % jfwang@hust.edu.cn
\affiliation{Wuhan National High Magnetic Field Center and School of Physics, Huazhong University of Science and Technology, Wuhan 430074, China}

\author{Zhangzhen He~(何长振)}  % hezz@fjirsm.ac.cn
\affiliation{State Key Laboratory of Structural Chemistry, Fujian Institute of Research on the Structure of Matter, Chinese Academy of Sciences, Fuzhou, Fujian 350002, China}

\author{Liang Li~(李亮)}   % liangli44@hust.edu.cn
\email{liangli44@hust.edu.cn}
\affiliation{Wuhan National High Magnetic Field Center and School of Physics, Huazhong University of Science and Technology, Wuhan 430074, China}
\affiliation{State Key Laboratory of Advanced Electromagnetic Technology, Huazhong University of Science and Technology, Wuhan 430074, China}

\author{Yongkang Luo (罗永康)~\orcidlink{0000-0002-6098-5767}}
\email{mpzslyk@gmail.com}
\affiliation{Wuhan National High Magnetic Field Center and School of Physics, Huazhong University of Science and Technology, Wuhan 430074, China}
\affiliation{State Key Laboratory of Advanced Electromagnetic Technology, Huazhong University of Science and Technology, Wuhan 430074, China}

\author{Jianda Wu (吴建达)~\orcidlink{0000-0002-3571-3348}}
\email{wujd@tongji.edu.cn}
\affiliation{School of Physics Science and Engineering, Tongji University, Shanghai 200092, China}

\begin{abstract}
We investigate the transverse-field thermodynamics of the
quasi-one-dimensional Ising-like antiferromagnet BaCo$_2$V$_2$O$_8$,
whose tilted screw-chain geometry and anisotropic Land\'e $g$ tensor
generate spatially modulated Zeeman couplings.
Angle-resolved magnetocaloric-effect (MCE) measurements reveal a high-field
temperature minimum near the transverse-field Ising critical field for
$H\parallel[110]$ that persists and shifts only weakly upon field rotation.
Tensor-network calculations show that the rotation-induced staggered transverse
field rapidly lowers the Ising critical field and that the magnetic Gr\"uneisen
ratio changes sign near the high-field temperature minimum, consistent with
experiment.
Our results establish that a dominant MCE response can
persist away from the Ising critical region,
suggesting a route to magnetic cooling by tailoring anisotropic
Zeeman-coupling configurations in quantum magnets.
\end{abstract}

\date{\today}
\maketitle
\end{CJK*}

\begin{figure}[t]
    \centering
    \panelgraphics[width=1.0\columnwidth]{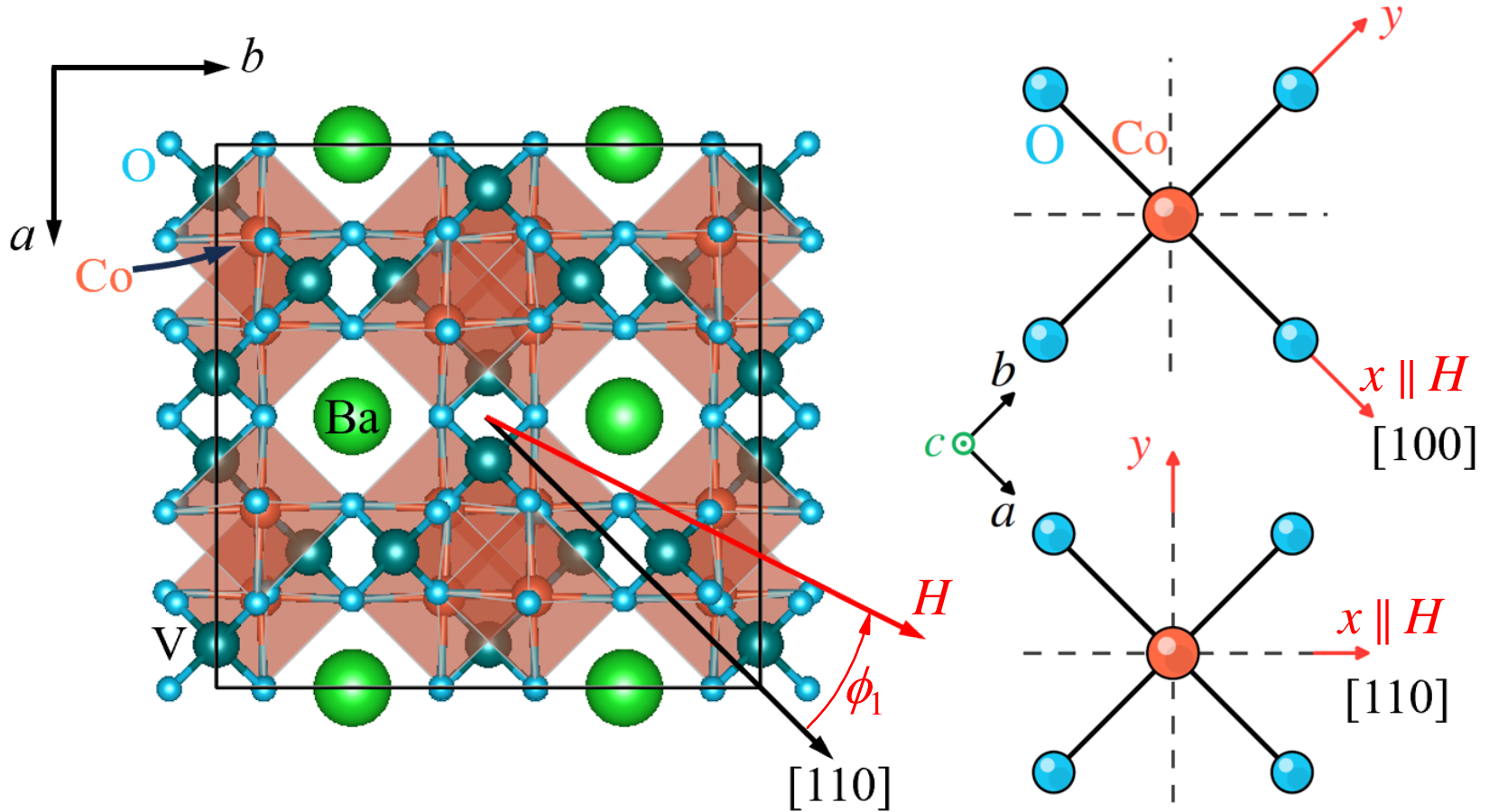}{(a)}{3}{92}
    \caption{Crystalline structure of BaCo$_2$V$_2$O$_8$ and geometry for the transverse-field directions $H\parallel[110]$ and $H\parallel[100]$ relative to the screw-chain local axes.
    }
    \label{fig:geo}
\end{figure}

\noindent{\it \bluefour{Introduction.---}}Low-dimensional quantum magnets offer experimentally
controllable platforms for studying quantum phase transitions.
A paradigmatic example is the transverse-field Ising chain (TFIC), which
provides direct access to quantum criticality, emergent conformal field theory, and universal scaling~\cite{sachdev2023QuantumPhasesMatter,pfeuty1970OnedimensionalIsingModel}.
The quasi-one-dimensional (1D) Ising-like antiferromagnets $A$Co$_2$V$_2$O$_8$ ($A=\mathrm{Ba},\mathrm{Sr}$) constitute
compelling material realizations of effective spin-$1/2$ Heisenberg-Ising (XXZ) chains that can be tuned by external magnetic fields~\cite{kimura2013CollapseMagneticOrder,yang2023MagneticExcitationsOnedimensional,zou2019UniversalityHeisenbergIsing,bera2017SpinonConfinementQuasionedimensional,fan2020PhaseDiagram,He2005CM,He2005PRB,Kimura2007PRL,Kimura2008a,Kimura2008b,Kawasaki2011,Faure2019,yang2022LocalDynamicsThermal,zhang2020ObservationE8,wu2018CrossoversCritical,wu2014FiniteTemperatureSpin}.
Owing to their tilted screw-chain structure and anisotropic Land\'{e} $g$ tensors in Fig.~\ref{fig:geo},
the materials exhibit non-negligible in-plane magnetic anisotropy, which
facilitates the realization of TFIC quantum criticality for fields applied along $[110]$ and $[100]$~\cite{cui2019QuantumCriticality,wang2018QuantumCriticality,faure2018TopologicalQuantumPhase}, exotic $E_8$ physics for $H\parallel[100]$~\cite{zou2021E8Spectra}, as well as many-body Bethe-string excitations in longitudinal field~\cite{wang2018ExperimentalObservation,yang2023ConfinementManybody,bera2020DispersionsManyBody,yang2024TruncatedString}.

Magnetocaloric effect (MCE), commonly quantified by the magnetic Gr\"uneisen ratio, provides a sensitive probe of field-induced quantum criticality and a basis for magnetic refrigeration~\cite{zhu2003UniversallyDiverging,garst2005SignChange,wolf2011Magnetocaloric,zhou2026UniversalScaling}.
Recent experiments have demonstrated giant MCEs associated with spin-supersolid physics in the insulating magnet Na$_2$BaCo(PO$_4$)$_2$ and the metallic magnet EuCo$_2$Al$_9$~\cite{xiang2024GiantMagnetocaloric,shu2026MetallicSpinSupersolid}.
In parallel, theory has proposed field tilting as a route to enhanced cooling in quantum-supercritical Ising magnets, highlighting field orientation as a versatile means of thermodynamic tuning~\cite{lv2025QuantumSupercritical}.
Yet in the $A$Co$_2$V$_2$O$_8$ Ising magnets, rotating the laboratory field within the $ab$ plane simultaneously changes several symmetry-distinct, spatially modulated Zeeman couplings because of the tilted screw-chain geometry and site-dependent anisotropic $g$ tensor~\cite{kimura2013CollapseMagneticOrder}.
This multicomponent coupling raises the central question of whether an angle-resolved MCE signal continues to track the intrinsic Ising critical field or is instead governed by transverse spin fluctuations.
Resolving this ambiguity is essential both for interpreting the MCE as a signature of quantum criticality and for guiding microscopic design principles for magnetic cooling.

In this Letter, we combine angle-resolved MCE and magnetization measurements with zero- and finite-temperature tensor-network calculations to resolve this issue in BaCo$_2$V$_2$O$_8$.
We show that field rotation separates the intrinsic Ising critical field from the dominant high-field magnetocaloric response.
The high-field response is governed mainly by the field-parallel transverse spin component.
These results imply a novel mechanism for magnetic cooling via engineering magnetic fluctuations through spin-component-selective redistribution of Zeeman couplings in quantum magnets.

\begin{figure}[t]
\centering
\includegraphics[width=\columnwidth]{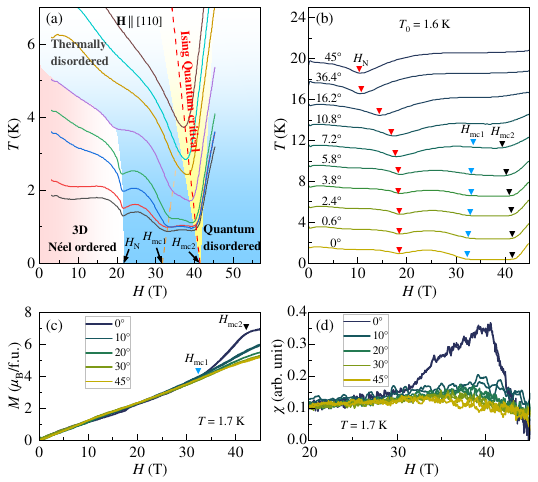}
\vspace*{-15pt}
\caption{Experimental MCE and magnetization data of BaCo$_2$V$_2$O$_8$ under magnetic field.
The angle $\phi_1=0^\circ$ denotes $H\parallel[110]$ and $\phi_1=45^\circ$ denotes $H\parallel[100]$.
(a) For $H\parallel[110]$, the temperature-dependent MCE data show that a single broad dip evolves into two dips ($H_{\mathrm{mc1}}$ and $H_{\mathrm{mc2}}$) as the temperature is lowered.
(b) The $T_0=1.6$ K MCE at various angles. The curves are vertically shifted for clarity.
(c) Angle-dependent magnetization ($M$) and (d) the corresponding magnetic susceptibility ($\chi=dM/dH$) with high-field features near $[110]$.
}
\label{fig:mce_M}
\end{figure}

\noindent{\it \bluefour{Experimental signatures.---}}High-field magnetocaloric and magnetization measurements up to 45 T were performed with the field rotated within the $ab$ plane at the Wuhan National High Magnetic Field Center, China [see Sec.~\ref{app:experimental_methods} of the SM~\cite{SM}].
We define $\phi_1$ as the in-plane angle measured from the $[110]$ direction [Fig.~\ref{fig:geo}].
For $H\parallel[110]$,
the data in Fig.~\ref{fig:mce_M}(a) show that,
upon cooling, a single dip evolves into two distinct dips
centered at $H_{\mathrm{mc1}}\approx32$ T and $H_{\mathrm{mc2}}\approx40$ T,
where the subscript $mc$ labels magnetocaloric effects.
$H_{\mathrm{mc2}}$ coincides
with the previously established 1D TFIC critical field
along [110]~\cite{cui2019QuantumCriticality,wang2018QuantumCriticality}.
The lower-field feature $H_{\mathrm{mc1}}$ is associated with a weaker anomaly observed in magnetization and,
although it may signal a first-order-like transition~\cite{kimura2013CollapseMagneticOrder}, is not the
main focus of this letter.

Fig.~\ref{fig:mce_M}(b) shows the angular evolution of the MCE for an initial temperature of $1.6$ K.
As the field is rotated away from $[110]$, the $H_{\mathrm{mc1}}$ dip rapidly weakens and disappears.
In contrast, the $H_{\mathrm{mc2}}$ dip remains the dominant thermodynamic feature
over a finite angular range, but vanishes for large $\phi_1$.
Both the temperature-dependent and angle-dependent MCE data
exhibit a clear low-field dip associated with the 3D N\'eel-order transition at $\HN$ in Figs.~\ref{fig:mce_M}(a,b)
[see Sec.~\ref{app:experimental_methods} of the SM~\cite{SM}].
$\HN$ shifts continuously from about $22$ T for $[110]$ to about $10$ T for $[100]$.
The magnetization data in Fig.~\ref{fig:mce_M}(c) show a clear kink near $H_{\mathrm{mc2}}\approx40$ T for $[110]$ with a shoulder around $H_{\mathrm{mc1}}\approx 32$ T.
The $H_{\mathrm{mc2}}$ feature weakens as the field is rotated and is absent for $[100]$ [Fig.~\ref{fig:mce_M}(d)], consistent with the MCE measurements.

The experimental observations in Fig.~\ref{fig:mce_M} therefore leave open whether the high-field MCE feature continues to track the intrinsic Ising transition as the field is rotated.
Alternatively, it may be governed by the response of a particular Zeeman-coupled spin component.
In the tilted screw-chain geometry, the anisotropic local $g$ tensor distributes the Zeeman coupling among several spin components.
Resolving their respective contributions requires a component-resolved microscopic calculation and a finite-temperature evaluation of the magnetic Gr\"uneisen ratio.
We therefore turn to an effective screw-chain model.

\noindent{\it \bluefour{Effective screw-chain model.---}}To investigate the microscopic origin of the experimental signatures,
we use an effective spin-$1/2$ XXZ chain with the site-dependent
Land\'e $g$ tensor imposed by the tilted screw-chain geometry~\cite{kimura2013CollapseMagneticOrder},
\begin{equation}
\begin{aligned}
\Heff
=&\,J\sum_j\left[
S_j^zS_{j+1}^z+
\epsilon\left(S_j^xS_{j+1}^x+S_j^yS_{j+1}^y\right)
\right]
\\
&+\mu_B H\sum_j\left[
g_{xx}^{(j)}S_j^x+g_{xy}^{(j)}S_j^y+g_{xz}^{(j)}S_j^z
\right],
\end{aligned}
\label{eq:heff}
\end{equation}
where $\epsilon=0.46$ is the exchange anisotropy used for BaCo$_2$V$_2$O$_8$~\cite{kimura2013CollapseMagneticOrder}.
The second term is the site-dependent Zeeman coupling to a laboratory field
of magnitude $H$, with the $x$ axis chosen parallel to the applied field.
The screw-chain structure enters through the four-site phase
[Sec.~\ref{app:g_geometry} of the Supplemental Material (SM)~\cite{SM}],
\begin{equation}
\phi_j=\phi_1+\frac{\pi}{2}(j-1), \qquad j=1,2,3,4
\end{equation}
which gives
$g_{xx}^{(j)}=(g_{\xi}\cos^2\theta+g_{\zeta}\sin^2\theta)
\cos^2\phi_j+g_{\psi}\sin^2\phi_j$,
$g_{xy}^{(j)}=\tfrac{1}{2}(g_{\xi}\cos^2\theta-g_{\psi}
+g_{\zeta}\sin^2\theta)\sin2\phi_j$, and
$g_{xz}^{(j)}=\tfrac{1}{2}(g_{\zeta}-g_{\xi})\sin2\theta\cos\phi_j$.
The anisotropic $g$ tensor resolves this Zeeman coupling into the
$g_{xx}^{(j)}S_j^x$, $g_{xy}^{(j)}S_j^y$, and $g_{xz}^{(j)}S_j^z$ channels,
each with a distinct angular dependence.
Since $\sin2\phi_j=(-1)^{j-1}\sin2\phi_1$, the $g_{xy}^{(j)}S_j^y$ channel is absent for $H\parallel[110]$ ($\phi_1=0^\circ$) and emerges upon field rotation as a staggered transverse field whose amplitude grows from zero as $\sin2\phi_1$~\cite{faure2021Solitonic}.
Previous works have observed the TFIC criticality for $H\parallel[110]$ and $[100]$~\cite{cui2019QuantumCriticality,wang2018QuantumCriticality}, but the intermediate angles have not been explored.
As shown below, the induced staggered transverse field rapidly shifts the Ising critical field and thereby separates the high-field MCE dip from the Ising transition.

At generic angles, the ordinary on-site spin-flip symmetry of the TFIC is absent. Nevertheless, the four-site field texture retains a combined symmetry $G_{2z}=T_2\Theta_z$, consisting of a two-site translation and an antiunitary spin operation that flips $S^z$ while leaving $S^{x,y}$ invariant (see Sec.~\ref{app:screw_symmetry} of the SM~\cite{SM}).

\begin{figure}[t]
\centering
\includegraphics[width=\columnwidth]{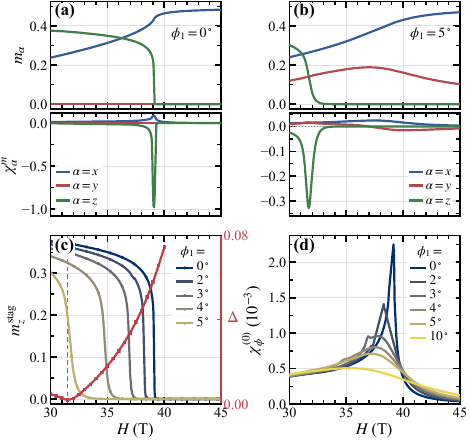}
\caption{
Magnetization, magnetic susceptibility and generalized susceptibility of the screw-chain XXZ model at zero-temperature.
(a), (b) Uniform transverse magnetization $m_x$, staggered components $m_y^{\rm stag}$ and $m_z^{\rm stag}$, and their field derivatives $\chi_{\alpha}^{m}$ for $\phi_1=0^\circ$ and $5^\circ$, respectively.
(c) Field dependence of $\msz$ for various $\phi_1$.
The red line shows the excitation gap $\Delta$ for $\phi_1=5^\circ$, whose minimum (vertical dashed line) follows the collapse of $m_z^{\rm stag}$ at $\HOne$.
(d) Generalized susceptibility $\chiphiZero$ for selected angles, showing the high-field response that defines $\HTwo$.
}
\label{fig:dmrg_fields}
\end{figure}

\noindent{\it \bluefour{Zero-temperature anisotropic response.---}}Based on the effective model in Eq.~\eqref{eq:heff}, we perform DMRG calculations~\cite{white1992DensityMatrix,fishman2022ITensor} of the uniform transverse magnetization $m_x$, the staggered components $m_y^{\rm stag}$ and $m_z^{\rm stag}$, and their field derivatives $\chi_{\alpha}^{m}$, with $\alpha=x,y,z$ (see Sec.~\ref{app:dmrg_details} of the SM~\cite{SM} for explicit definitions).
We identify the Ising critical field $\HOne$ from the collapse of the staggered-Ising order $\msz$, the associated extremum in $\chi_{z}^{m}$, and the excitation-gap minimum.
For $H\parallel[110]$, Fig.~\ref{fig:dmrg_fields}(a) shows that the staggered-Ising order $\msz$ collapses near $\HOne\approx 39$~T, where the transverse magnetization $m_x$ also saturates.
After a $5^\circ$ rotation away from $[110]$, $\HOne$ shifts to a lower field of $31.5$~T in Fig.~\ref{fig:dmrg_fields}(b), consistent with the gap minimum in Fig.~\ref{fig:dmrg_fields}(c).

The finite-temperature field response is characterized by the generalized
magnetization density and susceptibility derived from the free-energy density~\cite{zhu2003UniversallyDiverging,garst2005SignChange}:
\begin{equation}
\Mphi=-\frac{\partial f}{\partial H},\qquad
\chiphi=\frac{\partial \Mphi}{\partial H}=-\frac{\partial^2 f}{\partial H^2}.
\label{eq:chi_phi}
\end{equation}
Here the subscript $\phi$ denotes the response to variations in the field magnitude $H$ at fixed in-plane orientation $\phi_1$.
The quantity $f=F/N$ is the free-energy density, and $e_0=E_0/N$ is the
ground-state energy density. At zero temperature,
$f\to e_0$, and we use the superscript $(0)$ specifically for the
ground-state quantities,
$\MphiZero=-\partial e_0/\partial H$ and
$\chiphiZero=\partial\MphiZero/\partial H$.
We denote by $\HTwo$ the field of the prominent high-field maximum of $\chiphiZero$.
When the response becomes a broad hump, $\HTwo$ denotes the position of its maximum.
In Fig.~\ref{fig:dmrg_fields}(d),
the response defining $\HTwo$ remains near $37$--$39$~T for small rotations from $H\parallel[110]$ but broadens progressively with increasing $\phi_1$.
Beyond the angular range for which a discernible hump remains, $\HTwo$ can no longer be assigned robustly.

\begin{figure}[t]
\centering
\includegraphics[width=\columnwidth]{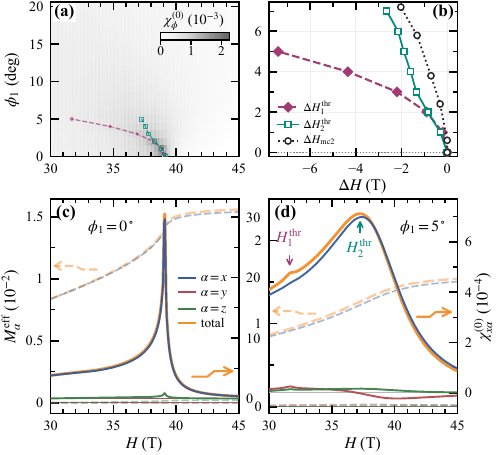}
\caption{
Angular evolution and component-resolved response of the screw-chain XXZ model at zero temperature.
(a) Intensity map of the generalized susceptibility $\chiphiZero$. Magenta diamonds and teal open squares mark $\HOne$ and $\HTwo$, respectively. (b) Angular shifts $\Delta H_i(\phi_1)=H_i(\phi_1)-H_i(0)$ of $\HOne$, $\HTwo$, and the experimental $H_{\mathrm{mc2}}$ values from Fig.~\ref{fig:mce_M}(b). Black open circles mark the experimental data.
(c),(d) Hellmann--Feynman decompositions of $\MphiZero$ (dashed curves, left axes) and $\chiphiZero$ (solid curves, right axes) for $\phi_1=0^\circ$ and $5^\circ$, respectively.
}
\label{fig:field_angle_hf}
\end{figure}

Fig.~\ref{fig:field_angle_hf}(a) displays the $\phi_1$--$H$ intensity map of $\chiphiZero$.
Fig.~\ref{fig:field_angle_hf}(b) summarizes the angular evolution of the two characteristic fields
and the experimental $H_{\mathrm{mc2}}$ using the field offsets
$\Delta H_i(\phi_1)=H_i(\phi_1)-H_i(0)$.
The weak angular shift of $H_{\mathrm{mc2}}$ is consistent with the high-field transverse response at $\HTwo$.
$\HOne$ and $\HTwo$ remain nearly degenerate for $0^\circ\leq\phi_1\lesssim2^\circ$.
Beyond this narrow range, $\HOne$ decreases rapidly to approximately $31.6$~T at $\phi_1=5^\circ$, whereas $\HTwo$ remains in the higher-field regime near $37$--$38$~T.
The resulting separation directly distinguishes the Ising transition at $\HOne$
from the generalized susceptibility response at $\HTwo$.

To clarify the microscopic origin of $\chiphiZero$, we apply the Hellmann--Feynman theorem~\cite{feynman1939ForcesMolecules} and
decompose the ground-state generalized magnetization $\MphiZero$, obtained from
the $T=0$ limit of Eq.~\eqref{eq:chi_phi}, into contributions from the
three local Zeeman channels,
\begin{equation}
\begin{aligned}
\MphiZero
&=-\frac{\mu_B}{N}\sum_j\left[
g_{xx}^{(j)}\langle S_j^x\rangle
+g_{xy}^{(j)}\langle S_j^y\rangle
+g_{xz}^{(j)}\langle S_j^z\rangle
\right]
\\
&\equiv\Mxe+\Mye+\Mze.
\end{aligned}
\label{eq:hf_mphi}
\end{equation}
Differentiating the three terms with respect to field gives
$\chi_{xx}^{(0)}$, $\chi_{xy}^{(0)}$, and $\chi_{xz}^{(0)}$, allowing us to identify the dominant response channel.

In Fig.~\ref{fig:field_angle_hf}(c,d), $\MphiZero$ is overwhelmingly dominated by $\Mxe$ for both $\phi_1=0^\circ$ and $\phi_1=5^\circ$, with only negligible contributions from $\Mze$ and $\Mye$.
Similarly, the peaks in $\chiphiZero$ are governed by the transverse channel $\chi_{xx}^{(0)}$, rather than by $\chi_{xy}^{(0)}$ or the longitudinal channel $\chi_{xz}^{(0)}$.
We therefore identify the field-parallel transverse Zeeman channel
$g_{xx}\langle S^x \rangle$ as the main contribution to the high-field response at $\HTwo$.
The $\chi_{xx}^{(0)}$ also exhibits a small side peak near $\HOne$ in Fig.~\ref{fig:field_angle_hf}(d), which is also visible for other small angles in Fig.~\ref{fig:dmrg_fields}(d).
The ground-state calculation therefore separates the Ising instability at
$\HOne$ from the dominant field-parallel transverse Zeeman response at $\HTwo$.

\noindent{\it \bluefour{Finite-temperature magnetocaloric response.---}}We next test this distinction at finite temperature by calculating the magnetic
Gr\"uneisen ratio of Eq.~\eqref{eq:heff},
\begin{equation}
\Gamma_H=\frac{1}{T}\left(\frac{\partial T}{\partial H}\right)_{S,\phi_1}
=-\frac{1}{C_{H,\phi_1}}
\left(\frac{\partial M_{\phi}}{\partial T}\right)_{H,\phi_1},
\label{eq:gamma_main}
\end{equation}
which is evaluated in the thermodynamic limit using
finite-temperature tensor-network methods~\cite{feiguin2005FiniteTemperatureDMRG,li2011LinearizedTensorRG,hauschild2018Efficient}.
Its numerator and denominator are evaluated, respectively, from the
energy--moment covariance and the energy variance of the same thermal state,
avoiding numerical differentiation with respect to temperature.
The detailed derivation and numerical implementation are given in
Sec.~\ref{app:mce_details} of the SM~\cite{SM}.
Because $C_{H,\phi_1}>0$,
an increasing field cools the spin system for
$\Gamma_H<0$ and heats it for $\Gamma_H>0$.
A negative-to-positive zero crossing therefore marks both an entropy maximum
at fixed temperature and a temperature minimum on an ideal isentrope, whereas
the positive peak measures the strength of the high-field MCE rather than, by
itself, a phase boundary~\cite{zhu2003UniversallyDiverging,garst2005SignChange,wolf2011Magnetocaloric}.

\begin{figure}[t]
    \centering
    \includegraphics[width=\columnwidth]{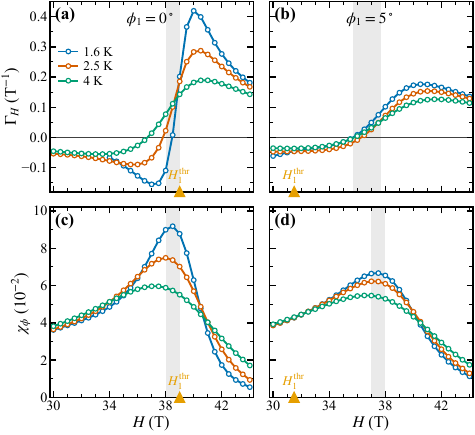}
    \caption{
    Finite-temperature Gr\"uneisen ratio $\Gamma_H$ (a,b) and susceptibility (c,d) of the screw-chain XXZ model for $\phi_1=0^\circ$
    (left) and $5^\circ$ (right).
    Gray bands mark the lowest-temperature $\Gamma_H$ sign-change regions and the corresponding lowest-temperature $\chi_{\phi}$ peak regions. Orange triangles mark the expected Ising transition fields $\HOne$.
    }
    \label{fig:mce_theory}
\end{figure}

For $\phi_1=0^\circ$, Fig.~\ref{fig:mce_theory}(a) displays a negative
low-field branch, a rapid sign change, and a positive high-field maximum that
returns gradually toward zero.
Upon lowering the temperature from $4$ to
$1.6$ K, the zero crossing moves from $36.7$ to $38.5$ T,
while the positive maximum shifts from $41.0$ to $40.0$ T.
Consequently, the field separation between the entropy maximum and the
positive MCE maximum decreases from $4.3$ T at $4$ K to $1.5$ T at $1.6$ K,
while the sign-changing feature becomes progressively sharper upon cooling.
This evolution occurs in the same field region where $\HOne$ and $\HTwo$
nearly coincide at zero temperature.

A $5^\circ$ rotation preserves the negative--positive trend
[Fig.~\ref{fig:mce_theory}(b)].
At $1.6$ K, the zero crossing occurs at $H_{\Gamma=0}=35.7$ T, while the broad
positive maximum remains at $40.5$ T and is reduced to $0.18$ T$^{-1}$.
The reduced amplitude and broad line shape contrast with the rapidly
sharpening $0^\circ$ response.

Figs.~\ref{fig:mce_theory}(c,d) show the
corresponding finite-temperature generalized susceptibility.  At $0^\circ$,
its peak sharpens on cooling and lies close to the low-temperature $\Gamma_H$
zero-crossing region.  At $5^\circ$, the peak remains broad in the high-field
regime and is clearly separated from $\HOne$.  The entropy maximum in
Fig.~\ref{fig:mce_theory}(b) lies below the susceptibility maximum in
Fig.~\ref{fig:mce_theory}(d) and the zero-temperature $\HTwo$.  This offset is
expected because these fields are extracted from different free-energy
derivatives at different temperatures.  Nevertheless, all three identify the
same high-field transverse-response regime and vary much more weakly with angle
than $\HOne$, which has already fallen to $31.5$ T.
This behavior is qualitatively consistent with the weak angular evolution of
the experimental $H_{\mathrm{mc2}}$ dip in Fig.~\ref{fig:mce_M}(b).

The sign reversal nevertheless indicates that an adiabatic field sweep toward
the zero crossing would cool the system from either side.  The sharper response
for $H\parallel[110]$ makes this orientation more favorable for cooling,
whereas a $5^\circ$ rotation weakens and broadens the effect.  Quantitative
cooling performance also depends on lattice and experimental heat
capacities and nonadiabatic losses, as discussed in
Sec.~\ref{app:mce_details} of the SM~\cite{SM}.

\noindent{\it \bluefour{Discussion and conclusions.---}}The screw symmetry and anisotropic Land\'e tensor allow $\HOne$ and $\HTwo$ to be distinguished as physically distinct fields.
Rotating the field away from $[110]$ activates the staggered $g_{xy}^{(j)}S_j^y$ term,
which preserves $G_{2z}$ and grows as $\sin2\phi_1$,
thereby destabilizing staggered $S^z$ order and rapidly lowering $\HOne$.
Nevertheless, $\HOne$ remains the Ising critical field, as indicated by the collapse of $m_z^{\rm stag}$ and the excitation-gap minimum.
For $H\parallel[110]$, $\HOne$ and $\HTwo$ nearly coincide near $39$ T,
so the experimental $H_{\mathrm{mc2}}$ dip in Fig.~\ref{fig:mce_M} probes the vicinity of the transverse-field Ising QCP~\cite{wang2018QuantumCriticality,wolf2011Magnetocaloric,zhou2026UniversalScaling}.
Upon field rotation, however, $\HOne$ shifts rapidly to lower fields,
while $\HTwo$ stays in the higher-field regime
and follows the broad generalized-susceptibility maximum
dominated by the field-parallel $g_{xx}^{(j)}\langle S_j^x\rangle$ channel.
Although $H_{\mathrm{mc2}}$, $H_{\Gamma=0}$, and $\HTwo$ are defined from different observables and need not coincide quantitatively,
their qualitative agreement, together with the Hellmann--Feynman decomposition,
associates the experimental $H_{\mathrm{mc2}}$ dip with the field-parallel transverse response represented by $\HTwo$, rather than with the collapse of the staggered $S^z$ order.

This interpretation places BaCo$_2$V$_2$O$_8$ in a regime distinct from established magnetic-refrigeration routes based on direct quantum-critical enhancement,
quantum-supercritical tuning by an order-parameter-conjugate field,
or supersolid fluctuations~\cite{zhu2003UniversallyDiverging,wolf2011Magnetocaloric,zhou2026UniversalScaling,lv2025QuantumSupercritical,xiang2024GiantMagnetocaloric,shu2026MetallicSpinSupersolid}.
In our case, the high-field response does not require invoking an additional ordered phase or a second quantum critical point at $\HTwo$.
Instead, the screw-chain geometry and anisotropic $g$ factors convert a single laboratory-field rotation into a redistribution among symmetry-distinct microscopic Zeeman channels, thereby separating the dominant magnetocaloric response from the intrinsic Ising transition.
This points to spin-channel selectivity as a possible design principle for magnetic cooling,
and suggests a broader route for engineering magnetocaloric responses through anisotropic Zeeman-coupling configurations in quantum magnets.

\noindent{\it \bluefour{Acknowledgments.---}}J.Y. thanks Xiao Wang for helpful discussions, particularly regarding the technical aspects of numerical calculations. C.D. acknowledges technical aids from Yoshimitsu Kohama.
This work is supported by National Key R\&D Program of China (2023YFA1609600, 2022YFA1602602).
J.W. is supported by the National Natural Science Foundation of China Nos. 12450004, 12274288, the Innovation Program for Quantum Science and Technology Grant No. 2021ZD0301900, and the Fundamental Research Funds for the Central Universities.
J.W. acknowledges the hospitality of Wilczek Quantum Center at Shanghai Institute for Advanced Studies of University of Science and Technology of China.
The high-field magnetization and magnetocaloric measurements were conducted at the Magnetic-property and Magnetocaloric Station of the Wuhan National High Magnetic Field Center (\url{https://whmfc.hust.edu.cn/yhfw/cljs.htm}).

\bibstyle{apsrev-nourl}
\bibliography{Refs}

\clearpage
\onecolumngrid
\raggedbottom

\begin{center}
\textbf{\large Supplemental Material for\\
``Anomalous magnetocaloric effects in the quasi-one-dimensional antiferromagnet BaCo$_2$V$_2$O$_8$''}
\end{center}

% \tableofcontents
\addtocontents{toc}{\protect\setcounter{tocdepth}{0}}
{
\tableofcontents
}

\setcounter{figure}{0}
\renewcommand{\thefigure}{S\arabic{figure}}
\renewcommand{\theHfigure}{S\arabic{figure}}
\setcounter{equation}{0}
\renewcommand{\theequation}{S\arabic{equation}}
\renewcommand{\theHequation}{S\arabic{equation}}
\setcounter{section}{0}
\renewcommand{\thesection}{\Roman{section}}
\setcounter{secnumdepth}{4}

\section{Pulsed-field magnetocaloric-effect and magnetization measurements}
\label{app:experimental_methods}

The magnetocaloric effect (MCE) and high-field magnetization were measured in pulsed magnetic fields up to 45~T at the Wuhan National High Magnetic Field Center.
For the MCE measurements, we followed established pulsed-field calorimetry and thermometry protocols~\cite{kohama2010ACMeasurement,kihara2013AdiabaticMeasurements}, using a directly deposited thin-film thermometer to obtain the required fast thermal response.
A $\sim100$-nm-thick AuGe film was deposited directly on the largest face of a BaCo$_2$V$_2$O$_8$ single crystal and patterned, together with Au contact pads, into a four-probe geometry; electrical leads were attached using silver paint.
The thermometer resistance was measured using an ac lock-in technique at 40~kHz, providing a sufficiently fast response to track the sample temperature during the field pulse.
The resistance was converted to temperature using its zero-field $R$--$T$ calibration, with a correction for the isothermal magnetoresistance of the AuGe film~\cite{kihara2013AdiabaticMeasurements}.
The slight difference between the $\HN$ values in Figs.~\ref{fig:mce_M}(a) and \ref{fig:mce_M}(b) arises from the different field-sweep rates used in the two measurements.

The high-field magnetization $M(H)$ was measured by the standard induction method using a coaxial pickup coil.
The background signal of the coil system was measured separately without the sample and subtracted from the sample signal.
The resulting $M(H)$ curves were calibrated against low-field magnetization data measured using a commercial SQUID magnetometer (Quantum Design).
The differential magnetic susceptibility $\mathrm{d}M/\mathrm{d}H$ shown in Fig.~\ref{fig:mce_M}(d) was obtained by numerical differentiation of the $M(H)$ curves.

\section{Geometry of the screw-chain Land\'e tensor}
\label{app:g_geometry}

The local $\xi\psi\zeta$ axes of the CoO$_6$ octahedra are tilted from the laboratory $xyz$ axes by an angle $\theta$ and rotate by $\pi/2$ from one Co site to the next along the screw chain.
In the laboratory frame the Land\'e tensor can be written as
\begin{equation}
\tilde g_{xyz}^{(j)}=
\begin{pmatrix}
g_{xx}^{(j)} & g_{xy}^{(j)} & g_{xz}^{(j)} \\
g_{xy}^{(j)} & g_{yy}^{(j)} & g_{yz}^{(j)} \\
g_{xz}^{(j)} & g_{yz}^{(j)} & g_{zz}^{(j)}
\end{pmatrix},
\end{equation}
with
\begin{equation}
\begin{aligned}
g_{xx}^{(j)} &=\left(g_{\xi}\cos^2\theta+g_{\zeta}\sin^2\theta\right)\cos^2\phi_j+g_{\psi}\sin^2\phi_j,\\
g_{yy}^{(j)} &=\left(g_{\xi}\cos^2\theta+g_{\zeta}\sin^2\theta\right)\sin^2\phi_j+g_{\psi}\cos^2\phi_j,\\
g_{zz}^{(j)} &=g_{\xi}\sin^2\theta+g_{\zeta}\cos^2\theta,\\
g_{xy}^{(j)} &=\frac{1}{2}\left(g_{\xi}\cos^2\theta-g_{\psi}+g_{\zeta}\sin^2\theta\right)\sin2\phi_j,\\
g_{yz}^{(j)} &=\frac{1}{2}\left(g_{\zeta}-g_{\xi}\right)\sin2\theta\sin\phi_j,\\
g_{xz}^{(j)} &=\frac{1}{2}\left(g_{\zeta}-g_{\xi}\right)\sin2\theta\cos\phi_j .
\end{aligned}
\end{equation}
For a transverse laboratory field along $x$, only the first column of this tensor enters the Zeeman coupling.
The site-dependent local phase is $\phi_j=\phi_1+(j-1)\pi/2$, where $\phi_1$ is the global field angle measured from $[110]$.

\begin{figure}[H]
\centering
\panelgraphics[width=0.42\textwidth]{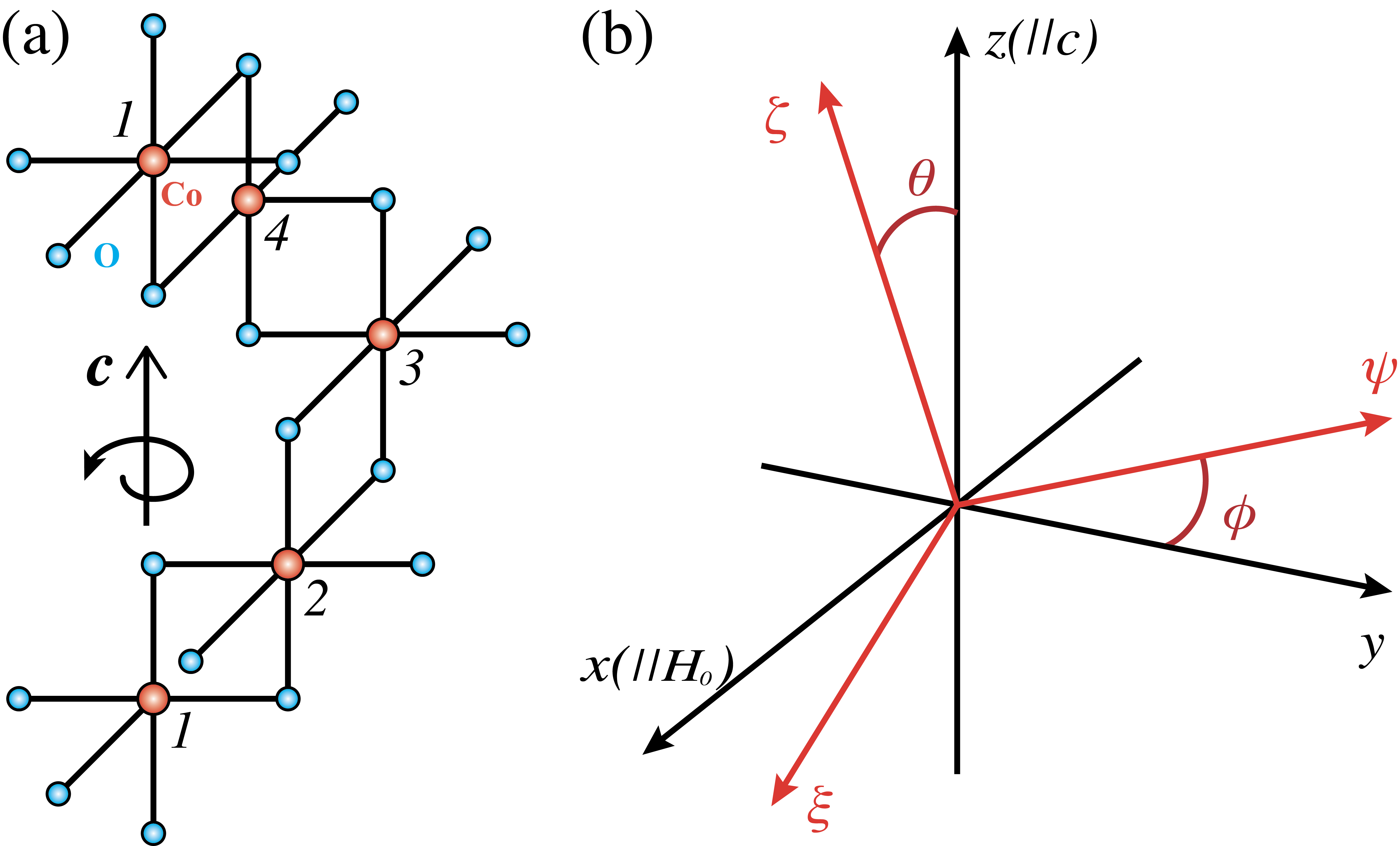}{}{3}{92}
\caption{
Screw-chain geometry and the relation between the laboratory and local coordinate systems.
}
\label{fig:sm_geometry}
\end{figure}

\section{Symmetry and field-amplitude analysis of the screw-chain effective Hamiltonian}
\label{app:screw_symmetry}

This section provides the symmetry basis for separating the two characteristic fields introduced in the main text.
In the numerical diagnostics used below, the order-parameter-collapse field $\Hz$ and the gap-minimum field $\Hgap$ identify $\HOne$, whereas the high-field maxima $\Hx$ of the field-parallel response and $\Hchi$ of the generalized susceptibility track $\HTwo$.
The starred quantities are auxiliary numerical diagnostics rather than additional characteristic fields.
We begin with the one-dimensional screw-chain Hamiltonian used in the main text,
\begin{align}
\mathcal H_{\rm eff}
=&\; J\sum_j\left[S_j^zS_{j+1}^z
+\epsilon\left(S_j^xS_{j+1}^x+S_j^yS_{j+1}^y\right)\right]
\nonumber\\
&+\mu_B H\sum_j\left[
 g_{xx}^{(j)}S_j^x+g_{xy}^{(j)}S_j^y+g_{xz}^{(j)}S_j^z
 \right] .
\label{eq:app_heff}
\end{align}
Here $z$ is the Ising axis, while the local $x$ axis is chosen parallel to the applied laboratory field within the basal plane.  The screw-chain structure enters through
\begin{equation}
\phi_j=\phi_1+\frac{\pi}{2}(j-1),
\label{eq:app_phij}
\end{equation}
where $\phi_1=0$ corresponds to $H\parallel[110]$ and $\phi_1=\pi/4$ corresponds to $H\parallel[100]$.  With
\begin{equation}
A\equiv g_\xi\cos^2\theta+g_\zeta\sin^2\theta,\qquad
B\equiv \frac{1}{2}(A-g_\psi),\qquad
C\equiv \frac{1}{2}(g_\zeta-g_\xi)\sin2\theta,
\label{eq:app_ABC}
\end{equation}
the field projections can be written as
\begin{align}
g_{xx}^{(j)} &= A\cos^2\phi_j+g_\psi\sin^2\phi_j,
\label{eq:app_gxx_raw}\\
g_{xy}^{(j)} &= B\sin2\phi_j,
\label{eq:app_gxy_raw}\\
g_{xz}^{(j)} &= C\cos\phi_j .
\label{eq:app_gxz_raw}
\end{align}
Equivalently, using Eq.~\eqref{eq:app_phij},
\begin{align}
g_{xx}^{(j)}
&=\bar g_x+\delta g_x(-1)^{j-1},
&
\bar g_x&=\frac{A+g_\psi}{2},
&
\delta g_x&=\frac{A-g_\psi}{2}\cos2\phi_1,
\label{eq:app_gxx_decompose}\\
g_{xy}^{(j)}
&=g_y^\pi(-1)^{j-1},
&
g_y^\pi&=\frac{A-g_\psi}{2}\sin2\phi_1,
\label{eq:app_gxy_decompose}\\
g_{xz}^{(j)}
&=C\cos\left[\phi_1+\frac{\pi}{2}(j-1)\right].
\label{eq:app_gxz_decompose}
\end{align}
Thus the $g_{xx}$ channel contains a uniform part and a two-sublattice modulation, the $g_{xy}$ channel is a staggered transverse field, and the $g_{xz}$ channel is a four-period longitudinal field.  These relations immediately imply
\begin{equation}
g_{xx}^{(j+2)}=g_{xx}^{(j)},\qquad
g_{xy}^{(j+2)}=g_{xy}^{(j)},\qquad
g_{xz}^{(j+2)}=-g_{xz}^{(j)} .
\label{eq:app_periodicity}
\end{equation}

\subsection{Combined screw symmetry at generic in-plane angles}
\label{app:generic_symmetry}

At a generic in-plane angle, the ordinary spin-flip symmetry of the ideal transverse-field Ising chain is absent because both induced channels $g_{xy}^{(j)}S_j^y$ and $g_{xz}^{(j)}S_j^z$ are finite.  Nevertheless, the screw-chain field texture in Eq.~\eqref{eq:app_periodicity} leaves a combined spin--lattice symmetry.  Define the antiunitary spin operation
\begin{equation}
\Theta_z=e^{-i\pi\sum_j S_j^z}\,\mathcal T,
\label{eq:app_Thetaz}
\end{equation}
where $\mathcal T$ is time reversal.  Its action on spin operators is
\begin{equation}
\Theta_z S_j^x\Theta_z^{-1}=S_j^x,
\qquad
\Theta_z S_j^y\Theta_z^{-1}=S_j^y,
\qquad
\Theta_z S_j^z\Theta_z^{-1}=-S_j^z .
\label{eq:app_Thetaz_action}
\end{equation}
Let $T_2$ denote translation by two sites along the chain.  The combined operation
\begin{equation}
G_{2z}\equiv T_2\Theta_z
\label{eq:app_G_generic}
\end{equation}
acts as
\begin{equation}
G_{2z} S_j^{x,y}G_{2z}^{-1}=S_{j+2}^{x,y},
\qquad
G_{2z} S_j^zG_{2z}^{-1}=-S_{j+2}^z .
\label{eq:app_G_action}
\end{equation}
The exchange part of Eq.~\eqref{eq:app_heff} is invariant under this operation.  The field part transforms as
\begin{align}
G_{2z}\mathcal H_{\rm field}G_{2z}^{-1}
=&+\mu_BH\sum_j\left[
 g_{xx}^{(j)}S_{j+2}^x+g_{xy}^{(j)}S_{j+2}^y-g_{xz}^{(j)}S_{j+2}^z
 \right]
\nonumber\\
=&+\mu_BH\sum_l\left[
 g_{xx}^{(l-2)}S_l^x+g_{xy}^{(l-2)}S_l^y-g_{xz}^{(l-2)}S_l^z
 \right]
\nonumber\\
=&+\mu_BH\sum_l\left[
 g_{xx}^{(l)}S_l^x+g_{xy}^{(l)}S_l^y+g_{xz}^{(l)}S_l^z
 \right]
=\mathcal H_{\rm field},
\label{eq:app_field_invariance}
\end{align}
where Eq.~\eqref{eq:app_periodicity} was used in the last step.  Thus $G_{2z}=T_2\Theta_z$ is an exact bulk symmetry of the ideal one-dimensional screw-chain Hamiltonian for any in-plane field angle. Since $G_{2z}^2$ is equivalent to a four-site translation up to the square of the spin operation, $G_{2z}$ supplies the $Z_2$-like operation that reverses the Ising order within the four-site screw unit cell.

This symmetry flips the staggered-Ising order parameter
\begin{equation}
m_z^{\rm stag}=\frac{1}{N}\sum_j(-1)^j\langle S_j^z\rangle .
\label{eq:app_mstag}
\end{equation}
Indeed,
\begin{equation}
G_{2z}\left[\sum_j(-1)^jS_j^z\right]G_{2z}^{-1}
=-\sum_j(-1)^jS_{j+2}^z
=-\sum_l(-1)^lS_l^z .
\label{eq:app_mstag_flip}
\end{equation}
Therefore the longitudinal projection $g_{xz}^{(j)}S_j^z$ does not act as a uniform conjugate field to $m_z^{\rm stag}$.  A true conjugate field to the staggered-Ising order would have the form
\begin{equation}
h_\pi^z\sum_j(-1)^jS_j^z,
\label{eq:app_conjugate_stag}
\end{equation}
which is odd under $G_{2z}$ and is therefore forbidden by the screw symmetry.  Equivalently, the four-period pattern $g_{xz}^{(j)}\propto\cos\phi_j$ has zero projection onto the staggered wave vector, as shown by
\begin{equation}
\sum_{j=1}^{4}(-1)^j\cos\left[\phi_1+\frac{\pi}{2}(j-1)\right]=0 .
\label{eq:app_zero_projection}
\end{equation}
The induced longitudinal projection can therefore shift or otherwise perturb the Ising critical point without immediately rounding it in the manner of a direct longitudinal field in the ideal TFIC.

\subsection{Special high-symmetry directions}
\label{app:special_directions}

The two experimentally important directions, $[110]$ and $[100]$, are special limits of the same field texture rather than distinct symmetry classes of the screw-chain model.

\paragraph*{$H\parallel[110]$.}
For $\phi_1=0$, one finds
\begin{equation}
\phi_j=0,\frac{\pi}{2},\pi,\frac{3\pi}{2},\ldots,
\label{eq:app_110_phis}
\end{equation}
and hence
\begin{align}
g_{xy}^{(j)}&=0,
\label{eq:app_110_gxy}\\
g_{xx}^{(j)}&=A,g_\psi,A,g_\psi,\ldots,
\label{eq:app_110_gxx}\\
g_{xz}^{(j)}&=C,0,-C,0,\ldots .
\label{eq:app_110_gxz}
\end{align}
Because the $g_{xy}$ channel is absent, the symmetry can be represented by the unitary operation
\begin{equation}
G_{[110]}=T_2P_x,
\qquad
P_x=\prod_j e^{-i\pi S_j^x},
\label{eq:app_G_110}
\end{equation}
under which $S^x$ is invariant while $S^{y,z}\to -S^{y,z}$.  The two-site translation changes $g_{xz}^{(j)}\to -g_{xz}^{(j)}$, which compensates the sign change of $S^z$.  If the weak four-period $g_{xz}^{(j)}S_j^z$ term is neglected, $G_{[110]}$ reduces to the ordinary TFIC spin-flip symmetry $P_x$.

\paragraph*{$H\parallel[100]$.}
For $\phi_1=\pi/4$, one obtains
\begin{equation}
\phi_j=\frac{\pi}{4},\frac{3\pi}{4},\frac{5\pi}{4},\frac{7\pi}{4},\ldots,
\label{eq:app_100_phis}
\end{equation}
which gives
\begin{align}
g_{xx}^{(j)}&=\frac{A+g_\psi}{2},
\label{eq:app_100_gxx}\\
g_{xy}^{(j)}&=\frac{A-g_\psi}{2}(-1)^{j-1},
\label{eq:app_100_gxy}\\
g_{xz}^{(j)}&=\frac{C}{\sqrt2}(+,-,-,+,\ldots).
\label{eq:app_100_gxz}
\end{align}
Thus $H\parallel[100]$ is characterized by a uniform field-parallel transverse component, a maximal staggered transverse component, and a weaker four-period longitudinal component.  It does not generally possess a simple unitary spin-only $Z_2$ symmetry such as $P_x$, because the staggered $g_{xy}^{(j)}S_j^y$ term is finite.  However, it still retains the combined antiunitary screw symmetry $G_{2z}=T_2\Theta_z$ discussed above.

\subsection{Field-amplitude hierarchy and physical consequences}
\label{app:field_amplitudes}

Equations~\eqref{eq:app_gxx_decompose}--\eqref{eq:app_gxz_decompose} show that the angular evolution of the effective fields is controlled by two simple trigonometric factors.  The staggered transverse field is
\begin{equation}
g_y^\pi=\frac{A-g_\psi}{2}\sin2\phi_1,
\label{eq:app_gy_angle}
\end{equation}
which vanishes for $H\parallel[110]$ and reaches its maximum magnitude for $H\parallel[100]$.  In contrast, the two-sublattice modulation of the $x$-field is
\begin{equation}
\delta g_x=\frac{A-g_\psi}{2}\cos2\phi_1,
\label{eq:app_gx_angle}
\end{equation}
which is maximal for $H\parallel[110]$ and vanishes for $H\parallel[100]$.  The longitudinal projection is controlled by
\begin{equation}
C=\frac{1}{2}(g_\zeta-g_\xi)\sin2\theta,
\label{eq:app_C_small}
\end{equation}
and is suppressed by the small tilt angle $\theta$ of the local octahedral axes.

For the parameter set used in the main text,
\begin{equation}
\theta=5^\circ,
\qquad
g_\xi=2.2,
\qquad
g_\psi=3.9,
\qquad
g_\zeta=6.2,
\label{eq:app_parameters}
\end{equation}
one obtains
\begin{equation}
A\simeq 2.23,
\qquad
\left|\frac{A-g_\psi}{2}\right|\simeq 0.84,
\qquad
|C|\simeq 0.35.
\label{eq:app_numeric_amplitudes}
\end{equation}
At $H\parallel[100]$, the actual four-period longitudinal amplitude is $|C|/\sqrt2\simeq 0.25$, whereas the staggered transverse amplitude is $|(A-g_\psi)/2|\simeq 0.84$.  Thus the induced staggered transverse channel is several times larger than the longitudinal projection.

This hierarchy has a direct physical consequence.  The $g_{xy}^{(j)}S_j^y$ term is transverse to the Ising axis and therefore enhances quantum fluctuations that suppress the staggered $S^z$ order.  As $\phi_1$ is increased away from $[110]$, this channel grows as $\sin2\phi_1$ and rapidly lowers the intrinsic Ising critical field.  By contrast, $g_{xz}^{(j)}S_j^z$ is a weaker four-period longitudinal perturbation with no staggered component at wave vector $\pi$; it perturbs the critical field only weakly and does not act as the conjugate field to $m_z^{\rm stag}$.  Consequently, $H\parallel[100]$ is best viewed not as a different symmetry class from generic in-plane angles, but as a staggered-transverse-field-dominated endpoint.  Small rotations away from $[110]$ instead define an intermediate regime in which the staggered transverse field is already strong enough to shift $\Hz$ downward, while the field-parallel transverse response can still remain dominated by the $g_{xx}S^x$ channel in the higher-field regime near $\Hx$ and $\Hchi$.

\section{DMRG implementation and numerical observables}
\label{app:dmrg_details}

\subsection{Model parameters and convergence criteria}

The DMRG calculations were implemented using ITensors.jl and ITensorMPS.jl~\cite{fishman2022ITensor} for the easy-axis XXZ chain in Eq.~\eqref{eq:app_heff} with $\epsilon=0.46$ and open boundary conditions.
The energy unit is $J=65$ K $=5.6$ meV, and magnetic fields are converted to tesla using $\mu_B=5.8\times10^{-2}$ meV/T.
The local tensor parameters are $\theta=5^\circ$, $g_{\xi}=2.2$, $g_{\psi}=3.9$, and $g_{\zeta}=6.2$.
Unless otherwise stated, the results in the main figures were obtained for $N=800$ sites.

For each field orientation, the first field point was initialized with a random matrix-product state, and subsequent field points were warm-started from the converged state at the preceding field.
We performed 200 DMRG sweeps, with the maximum allowed bond dimension set to 1000 for the first 100 sweeps and 2000 thereafter.
The singular-value cutoff and noise amplitude were both $10^{-10}$.
For every full sweep $s$, we recorded the energy and the maximum two-site truncation error over all bonds.
A state was accepted as converged when
\begin{equation}
\frac{|E^{(s)}-E^{(s-1)}|}{N}<10^{-10}J,
\qquad
\varepsilon_{\rm tr}^{(s)}<10^{-9}.
\label{eq:app_dmrg_convergence}
\end{equation}
All ground-state points displayed in the main figures satisfy these criteria.
The largest bond dimension retained in the reported $N=800$ ground states is 51, well below the imposed upper bound.

\subsection{Bulk magnetizations and response derivatives}
\label{app:dmrg_observables}

To reduce boundary effects in local observables, we use the central half of the open chain,
\begin{equation}
\mathcal B=\left\{\frac{N}{4}+1,\ldots,\frac{3N}{4}\right\},
\qquad N_{\mathcal B}=\frac{N}{2}.
\label{eq:app_bulk_window}
\end{equation}
For $N=800$, this corresponds to sites 201--600.
The uniform field-parallel magnetization and the staggered transverse and Ising components are defined as
\begin{align}
m_x
&=\left|\frac{1}{N_{\mathcal B}}
\sum_{j\in\mathcal B}\langle S_j^x\rangle\right|,
\label{eq:app_mx_bulk}\\
m_{\alpha}^{\rm stag}
&=\left|\frac{1}{N_{\mathcal B}}
\sum_{j\in\mathcal B}(-1)^j\langle S_j^{\alpha}\rangle\right|,
\qquad \alpha=y,z.
\label{eq:app_mstag_bulk}
\end{align}
Their signed field derivatives are
\begin{equation}
\chi_x^m=\frac{\partial m_x}{\partial H},
\qquad
\chi_{\alpha}^m=\frac{\partial m_{\alpha}^{\rm stag}}{\partial H},
\quad \alpha=y,z.
\label{eq:app_chim_def}
\end{equation}
Because $m_z^{\rm stag}$ decreases with increasing field, $\chi_z^m$ has a negative extremum; $\Hz$ is therefore determined from the maximum of $|\chi_z^m|$.
The energy density $e_0=E_0/N$ and the excitation energies are calculated from the full open chain, whereas the order parameters and the component-resolved Hellmann--Feynman expectation values use the bulk window in Eq.~\eqref{eq:app_bulk_window}.

For the Hamiltonian convention in Eq.~\eqref{eq:app_heff}, the field-conjugate operator and the generalized magnetization are
\begin{equation}
V_{\phi}=\frac{\partial\Heff}{\partial H},
\qquad
M_{\phi}^{(0)}=-\frac{\langle V_{\phi}\rangle}{N}
=-\frac{\partial e_0}{\partial H}.
\label{eq:app_hf_operator}
\end{equation}
Consequently,
\begin{equation}
\chi_{\phi}^{(0)}
=\frac{\partial M_{\phi}^{(0)}}{\partial H}
=-\frac{\partial^2e_0}{\partial H^2}
=-\frac{\partial}{\partial H}
\frac{\langle V_{\phi}\rangle}{N}.
\label{eq:app_chiphi_def}
\end{equation}
The last expression is used for the Hellmann--Feynman decomposition into the $g_{xx}S^x$, $g_{xy}S^y$, and $g_{xz}S^z$ channels.
The near-$[110]$ ground-state scans used a field spacing of $0.05$ T, with additional points spaced by $0.01$--$0.02$ T around the $\phi_1=5^\circ$ order-parameter collapse; broader-angle survey curves used a spacing of $0.5$ T.

\subsection{Excitation gap}

Since the transverse field terms do not conserve $S_{\rm tot}^z$, the lowest excitation was obtained by orthogonality-constrained excited-state DMRG rather than by optimizing in a fixed magnetization sector.
After obtaining the ground state $|\psi_0\rangle$, the first excited state was found by minimizing the penalized Hamiltonian
\begin{equation}
\widetilde{\mathcal H}_1
=\mathcal H_{\rm eff}
+\lambda|\psi_0\rangle\langle\psi_0|,
\qquad \lambda=20J,
\label{eq:app_excited_penalty}
\end{equation}
using the same sweep, bond-dimension, truncation, and energy-convergence settings as for the ground state.
The finite-size excitation gap is defined as
\begin{equation}
\Delta_N(H,\phi_1)=E_1(H,\phi_1)-E_0(H,\phi_1),
\qquad
\Hgap=\operatorname*{arg\,min}_{H}\Delta_N(H,\phi_1).
\label{eq:app_gap_def}
\end{equation}
Only the lowest excited state was targeted for the $N=800$, $\phi_1=5^\circ$ gap curve shown in Fig.~\ref{fig:dmrg_fields}(c); higher excited states are not used in determining $\Hgap$.
The gap curve was calculated on a $0.5$ T field grid.
At its minimum, $H=31.5$ T, both states satisfy Eq.~\eqref{eq:app_dmrg_convergence} and the residual overlap is $|\langle\psi_0|\psi_1\rangle|=3.7\times10^{-8}$.
Because $N$ is finite, the quantity displayed in the main text is a gap minimum rather than an exact zero of $\Delta_N$.

\section{Finite-temperature magnetocaloric calculation}
\label{app:mce_details}

\subsection{Thermodynamic identities and covariance estimator}

For a fixed in-plane field orientation, the magnetic Gr\"uneisen ratio probes
the field derivative of the entropy.  We use thermodynamic densities
$f=F/N$, $s=S/N$, and
$C_{H,\phi_1}=T(\partial s/\partial T)_{H,\phi_1}$, together with the
generalized magnetization density $M_{\phi}=-\partial f/\partial H$.  Then
\begin{equation}
\Gamma_H=
\frac{1}{T}\left(\frac{\partial T}{\partial H}\right)_{S,\phi_1}
=-\frac{1}{C_{H,\phi_1}}
\left(\frac{\partial s}{\partial H}\right)_{T,\phi_1}
=-\frac{1}{C_{H,\phi_1}}
\left(\frac{\partial M_{\phi}}{\partial T}\right)_{H,\phi_1},
\label{eq:app_gamma_thermo}
\end{equation}
where the last equality follows from the Maxwell relation.  Using $S$ rather
than $s$ in the isentropic derivative is equivalent because $N$ is fixed.

The energy--moment covariance used numerically follows directly from the
canonical ensemble~\cite{trippe2010ExactMagnetocaloric,yu2020GruneisenParameters}.  Let
\begin{equation}
\widehat{\mathcal M}_{\phi}=-\frac{\partial\Heff}{\partial H},
\qquad
\beta=\frac{1}{k_BT},
\qquad
Z={\rm Tr}\,e^{-\beta\Heff}.
\label{eq:app_moment_operator}
\end{equation}
Here $\widehat{\mathcal M}_{\phi}$ is the extensive field-conjugate operator,
while $M_{\phi}=\langle\widehat{\mathcal M}_{\phi}\rangle/N$ is its density.
For operators without explicit temperature dependence,
\begin{align}
C_{H,\phi_1}
&=\frac{k_B\beta^2}{N}\operatorname{Var}(\Heff),
\nonumber\\
-\left(\frac{\partial M_{\phi}}{\partial T}\right)_{H,\phi_1}
&=-\frac{k_B\beta^2}{N}\operatorname{Cov}
\left(\widehat{\mathcal M}_{\phi},\Heff\right),
\label{eq:app_covariance_thermo}
\end{align}
where $\operatorname{Cov}(A,B)=\langle AB\rangle-\langle A\rangle
\langle B\rangle$.  The common factors and the system size cancel, giving
the derivative-free estimator
\begin{equation}
\boxed{
\Gamma_H=-\frac{\operatorname{Re}\operatorname{Cov}
(\widehat{\mathcal M}_{\phi},\Heff)}
{\operatorname{Var}(\Heff)}} .
\label{eq:app_gamma_covariance}
\end{equation}
The covariance is real in an exact thermal state; its residual imaginary part
in the tensor-network contraction is retained as an independent numerical
diagnostic.

Equation~\eqref{eq:app_gamma_covariance} also makes the physical content
transparent.  In the energy eigenbasis,
\begin{equation}
M_{\phi}=\sum_n p_n\mathcal M_{n,\phi},
\quad
p_n=\frac{e^{-\beta E_n}}{Z},
\quad
\mathcal M_{n,\phi}=-\frac{1}{N}\frac{\partial E_n}{\partial H},
\label{eq:app_gamma_spectral}
\end{equation}
and hence
\begin{equation}
\left(\frac{\partial M_{\phi}}{\partial T}\right)_{H,\phi_1}
=k_B\beta^2
\left[\langle E\mathcal M_{\phi}\rangle
-\langle E\rangle\langle\mathcal M_{\phi}\rangle\right].
\label{eq:app_spectral_derivative}
\end{equation}
Thus an MCE anomaly measures a redistribution of thermal weight among
many-body states carrying different field-conjugate moments; it need not
coincide with the collapse of a particular order parameter.

For the infinite chain, we evaluate the extensive covariances as sums of
connected local correlators.  We assign the Zeeman term on site $a$ and the
exchange bond to its right to a local energy density $h_a$, and define
$m_a=-\partial h_a/\partial H$.  With the four-site screw unit cell,
\begin{align}
v_H(R)&=\frac{1}{4}\sum_{a=0}^{3}\sum_{r=-R}^{R}
\langle\delta h_a\,\delta h_{a+r}\rangle,
\nonumber\\
c_{MH}(R)&=\frac{1}{4}\sum_{a=0}^{3}\sum_{r=-R}^{R}
\langle\delta m_a\,\delta h_{a+r}\rangle,
\label{eq:app_local_covariance_sums}
\end{align}
where $\delta O=O-\langle O\rangle$.  Averaging all four origins and both
positive and negative displacements gives
\begin{equation}
C_{H,\phi_1}=k_B\beta^2v_H,
\qquad
\Gamma_H=-\frac{\operatorname{Re}c_{MH}}{v_H}.
\label{eq:app_local_gamma}
\end{equation}
The local operators are centered before contraction, which avoids subtracting
two large disconnected contributions after the correlation sum.

\subsection{Infinite-MPS purification and production grid}

The thermal density matrix is represented by purifying each physical spin
with an ancilla spin~\cite{feiguin2005FiniteTemperatureDMRG}.  Starting from
the maximally entangled infinite-temperature state $|\Psi(0)\rangle$, we form
\begin{equation}
|\Psi(\beta)\rangle\propto
\left(e^{-\beta\Heff/2}\otimes\mathbbm{1}_{\rm anc}\right)|\Psi(0)\rangle,
\qquad
\langle O\rangle_\beta=\frac{\langle\Psi(\beta)|O|\Psi(\beta)\rangle}
{\langle\Psi(\beta)|\Psi(\beta)\rangle}.
\label{eq:app_purification}
\end{equation}
We use a translationally invariant purification MPS with the physical
four-site screw unit cell and no conserved spin quantum number.  Imaginary
time is evolved by second-order TEBD as implemented with TeNPy~\cite{hauschild2018Efficient},
with ket step $d\tau=0.005$ meV$^{-1}$; because the ket carries
$e^{-\beta\Heff/2}$, the physical inverse-temperature increment is
$d\beta=2d\tau$.  The final step is aligned exactly to each target $\beta$.

The finite-temperature Hamiltonian parameters are identical to those in the
ground-state calculation: $J=5.6$ meV, $\epsilon=0.46$,
$\mu_B=0.057883817982$ meV/T, $\theta=5^\circ$, and
$(g_\xi,g_\psi,g_\zeta)=(2.2,3.9,6.2)$.  The production grid contains
$\phi_1=0^\circ$ and $5^\circ$, $T=1.6$, $2.5$, and $4.0$ K, and
$H=30$--$44$ T in $0.5$ T steps.  We impose the bond-dimension cap
$\chi_{\max}=128$, singular-value cutoff $10^{-13}$, and zero discarded-weight
cutoff beyond the bond-dimension truncation.

At each thermal state, Eq.~\eqref{eq:app_local_covariance_sums} is evaluated
for a cutoff ladder $R=128,256,512$.  Difficult low-temperature points are
extended to $R=768$ and $1024$.  A point is accepted only if the last two
cutoffs obey
\begin{equation}
\frac{|\Delta v_H|}{|v_H|}\leq10^{-3},
\qquad
|\Delta\Gamma_H|\leq10^{-4}\ {\rm T}^{-1},
\qquad
v_H>10^{-12}\ {\rm meV}^2,
\label{eq:app_covariance_acceptance}
\end{equation}
and the relative imaginary residuals of both covariance sums are below
$10^{-8}$.  The relative change of $c_{MH}$ is recorded but is not used as a
hard criterion because it is ill-conditioned at a physical zero crossing.
Altogether 166 points pass these tests and enter Fig.~\ref{fig:mce_theory}.
The only omitted interval is $\phi_1=0^\circ$, $T=1.6$ K,
$H=30$--$33.5$ T, for which the small variance was not cutoff-resolved even at
$R=1024$; the displayed $1.6$ K curve therefore begins at $34$ T.

\begin{table}[H]
\caption{Features extracted directly from the resolved $0.5$ T production
grid.  $H_{\Gamma=0}$ is obtained by linear interpolation across the sign
change and $H_{\rm pk}$ is the positive grid maximum.  Interpolated fields
should not be assigned an accuracy better than the $\pm0.25$ T geometric
half-step.}
\label{tab:app_mce_features}
\begin{ruledtabular}
\begin{tabular}{cccc}
$\phi_1$ & $T$ (K) & $H_{\Gamma=0}$ (T) &
$H_{\rm pk}$ (T), $\Gamma_{H,\rm pk}$ (T$^{-1}$)\\
\hline
$0^\circ$ & 1.6 & 38.5 & 40.0, 0.42\\
$0^\circ$ & 2.5 & 37.8 & 40.5, 0.29\\
$0^\circ$ & 4.0 & 36.7 & 41.0, 0.19\\
$5^\circ$ & 1.6 & 35.7 & 40.5, 0.18\\
$5^\circ$ & 2.5 & 36.3 & 41.0, 0.15\\
$5^\circ$ & 4.0 & 35.8 & 41.5, 0.13\\
\end{tabular}
\end{ruledtabular}
\end{table}

\subsection{Relation to measured $T(H)$ and field-channel selectivity}
\label{app:mce_experiment}

For an ideal isentropic sweep, Eq.~\eqref{eq:app_gamma_thermo} implies
\begin{equation}
\frac{d\ln T}{dH}=\Gamma_H[T(H),H].
\label{eq:app_isentrope_ode}
\end{equation}
Therefore a temperature minimum on an ideal $T(H)$ path occurs where
$\Gamma_H[T(H),H]=0$.  The positive maximum instead quantifies the strongest
local logarithmic heating response at fixed temperature and must not be
identified as the isentropic minimum itself.  On the negative side of the zero
crossing an up-sweep cools the system, while on the positive side a down-sweep
cools it, so an adiabatic field sweep toward the entropy ridge cools from
either direction.  Reconstructing a particular experimental path would require
evaluating $\Gamma_H$ along the temperatures actually reached by that path.

The calculation contains the magnetic single-chain heat-capacity density only.
If a field-independent background heat-capacity density $C_{\rm bg}$ from
phonons and the experimental setup is included while the magnetic numerator is
unchanged,
the observed ratio becomes
\begin{equation}
\Gamma_H^{\rm obs}\simeq
-\frac{\bigl(\partial M_{\phi}/\partial T\bigr)_{H,\phi_1}^{\rm mag}}
{C_{H,\phi_1}^{\rm mag}+C_{\rm bg}}.
\label{eq:app_gamma_background}
\end{equation}
Such a background suppresses the absolute amplitude but does not by itself
move the sign change.  Heat leaks, sweep-rate effects, and additional
field-dependent degrees of freedom can modify the measured path further.
Accordingly, the field evolution and angle-induced broadening in
Fig.~\ref{fig:mce_theory} provide the more robust comparison to experiment
than its absolute single-chain amplitude.

An MCE feature can arise from a phase boundary, a gap closing or minimum, or a
crossover among thermally occupied states with different
$\mathcal M_{n,\phi}$.  Conversely, a microscopic instability can be weak or
unresolved when it carries little entropy in the thermal window or projects
weakly onto $M_{\phi}$.  This distinction is central in BCVO: the
rotation-induced staggered $g_{xy}^{(j)}S_j^y$ field enhances transverse
mixing and rapidly lowers the Ising field, whereas the field-conjugate
$g_{xx}^{(j)}S_j^x$ channel dominates the higher-field generalized response.
The persistence of a broadened finite-temperature MCE response near this
transverse-response field, after the Ising field has moved to lower field, is therefore
consistent with fluctuation-assisted cooling and does not require assigning a
second quantum critical point to the high-field feature.

\end{document}